\documentstyle[twocolumn,floats,amssymb,aps,epsfig,prl]{revtex}
\newcommand{\lb}{{\langle}} \newcommand{\rb}{{\rangle}}

\begin{document}
\draft 
\wideabs{ 

\title{Clustering transitions in vibro-fluidized
magnetized granular materials} 
\author{Daniel L. Blair and A. Kudrolli}  
\address{Department of Physics, Clark University, Worcester, MA 01610,
USA}  
\date{\today} 
\maketitle

\begin{abstract}
We study the effects of long range interactions on the
phases observed in cohesive granular materials.  At high vibration
amplitudes, a gas of magnetized particles is observed with velocity
distributions similar to non-magnetized particles. Below a transition
temperature compact clusters are observed to form and coexist with
single particles. The cluster growth rate is consistent with a
classical nucleation process. However, the temperature of the
particles in the clusters is significantly lower than
the surrounding gas, indicating a breakdown of equipartition. If the
system is quenched to low temperatures, a meta-stable network of
connected chains self-assemble due to the anisotropic nature of
magnetic interactions between particles.
\end{abstract}
}

Vibro-fluidized inelastic spherical particles are an important model
system which capture the essence of dissipative interactions on the
statistical properties of dry non-cohesive granular
materials~\cite{jaeger96}.  This
system has emerged as an important test bed to investigate the
applicability of dissipative kinetic
theory~\cite{haff-jenkins83,simulations}. Experiments measuring the
position and velocity of individual particles show the formation of
clusters, non-Gaussian velocity distributions (to varying extent), and
the violation of equipartition~\cite{los-olafsen99,krb}. In a number
of applications, additional cohesive interactions often exist due to
the presence of moisture, electrostatic screening, and
magnetization~\cite{cohesive}.

In this Letter, we introduce a novel system consisting of magnetized
particles inside a vibrated container. This enables us to study the
effect of long-range attractive interactions on the formation of
clusters and the velocity distributions of cohesive granular
materials. Non-magnetic particles are also present in our system to
define a system temperature that depends on the vibration amplitude of
the container. A gas-like phase of magnetized particles is observed at
high system temperature. If the system is slowly cooled below a
transition temperature, compact clusters precipitate and grow in time.
If the temperature is rapidly quenched, an extended mesh of particles
is observed to self-assemble due to the anisotropy of the
interaction. The structure and growth of the clusters is consistent
with both a classical nucleation process, and recent theories of the
dipolar hard sphere
model~\cite{leeuwen93,sear96,camp00,tlusty00}. However, the velocity
distributions and the associated granular temperature of the free and
clustered particles show the influence of dissipation and cohesion.

The idealized interaction between two dipolar hard spheres separated
by distance $r$ is defined as
\begin{equation}
U = U_{HS} + \frac{1}{r^3}  (\vec{\mu_i} \cdot \vec{\mu_j}) -
\frac{3}{r^5} (\vec{\mu_i} \cdot \vec{r}_{ij})(\vec{\mu_j} \cdot
\vec{r}_{ij}),
\label{eq:dhs}
\end{equation}
where $U_{HS}$ corresponds to the hard core repulsion interaction,
$\vec{\mu}$ is the dipole moment, and $\vec{r}$ is the  interparticle
vector connecting the centers of dipoles $i,j$. This type of
interaction has been put forth as an extremum model for
ferro-fluids~\cite{rosensweig,degen_pinc}.  If the potential is
averaged over all possible dipole arrangements, the familiar $r^{-6}$
van der Waals interaction is recovered.  This implies that such
systems should have a well defined liquid gas transition.  However,
extensive simulation and theoretical work on the dipolar hard sphere
model have not conclusively determined the existence of critical phase
transition.  Recent experiments on ferro-colloid suspensions suggest
the existence of a liquid gas transition, but  open questions still
remain~\cite{mamiya00}.  Other experiments using nickel coated glass
micro-spheres in a liquid solution have found the existence of ring
and mesh-like states~\cite{wen}.

The particles used in our experiments are chrome steel spheres with a
diameter of $\sigma = 0.3$ cm (with a high degree of sphericity
$\delta \sigma/\sigma \sim 10^{-4}$) and mass $m = 0.12$~g.  Each
sphere has been placed in a ramped field of $1 \times 10^4$~G to embed
a permanent moment of $\mu \sim 10^{-2}$~emu per particle. The
apparatus consists of a 30.0 cm diameter flat, anodized aluminum
plate, with 1.0~cm sidewalls and a clear acrylic lid,  that is weakly
coupled through a rigid linear bearing to an electromechanical shaker.
The system is leveled to within 0.001 cm to ensure that the plate is
uniformly accelerated. The measured acceleration of the plate $\Gamma
= {\cal A}\omega^2 / g$, where $\cal{A}, \omega$ are the amplitude and
angular frequency and $g$ is the acceleration due to gravity, is
varied between $\Gamma = 0- 3.0 \,g$, at  $\omega = 377$ rad
s$^{-1}$. Image data is acquired through a high speed Kodak SR-1000
digital camera with a spatial (temporal) resolution of 512 $\times$
480 pixels (250 f.p.s.).  The covering fraction of magnetic particles,
$\phi $, defined as the ratio of the area of the particles to that of
a close packed mono-layer, is varied  from $\phi~=~0.01~\to~0.15$.  We
also place glass particles of equivalent mass, with a fixed volume
fraction of $\phi_{p} = 0.15$ to introduce additional stochasticity
into the system.   The glass particles act as thermal carriers to
ensure that as the magnetic particles condense, the temperature of the
system does not go to zero.

The granular temperature is defined as~$T_g = \frac{1}{2}m\lb{\bf
v}^2\rb$.  By measuring  velocity components along the two horizontal
axes ${\bf v}_{x,y}$ we build a distribution of velocities whose
width, given by the second moment, determines the granular
temperature. The form of the  ${\bf v}_{x,y}$-distributions deviate
from a Gaussian consistent with previous
observations~\cite{los-olafsen99}.  The system temperature $T$, is
determined by the velocity distribution of the gas of both magnetic
and non-magnetic particles.   We will return to a more detailed
analysis of the velocity distributions for the gas and clustered
phases after discussing the nature of the clusters.

\begin{figure}[t]
\centerline{\epsfig{file=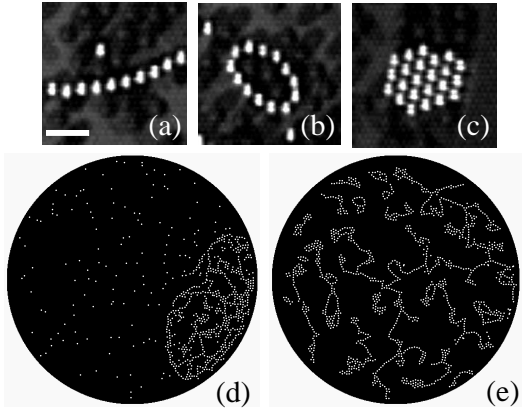,width=7.0cm}}
\caption{Examples of chains (a), rings (b), and crystallites (c)
observed. Rings appear be the most stable
configuration. The scale bar denotes 1~cm.  (d) Snapshot of the system
at $\phi=0.09$, where $T$ is lowered to $T_s$ from the gas state after
1092~s.  (c) The system at $\phi = 0.15$ after a rapid quench from the
gas state into the network state.  Images (c,d) are the size of the
system.}
\label{fig:images}
\end{figure}
\begin{figure}[t]
\centerline{\epsfig{file=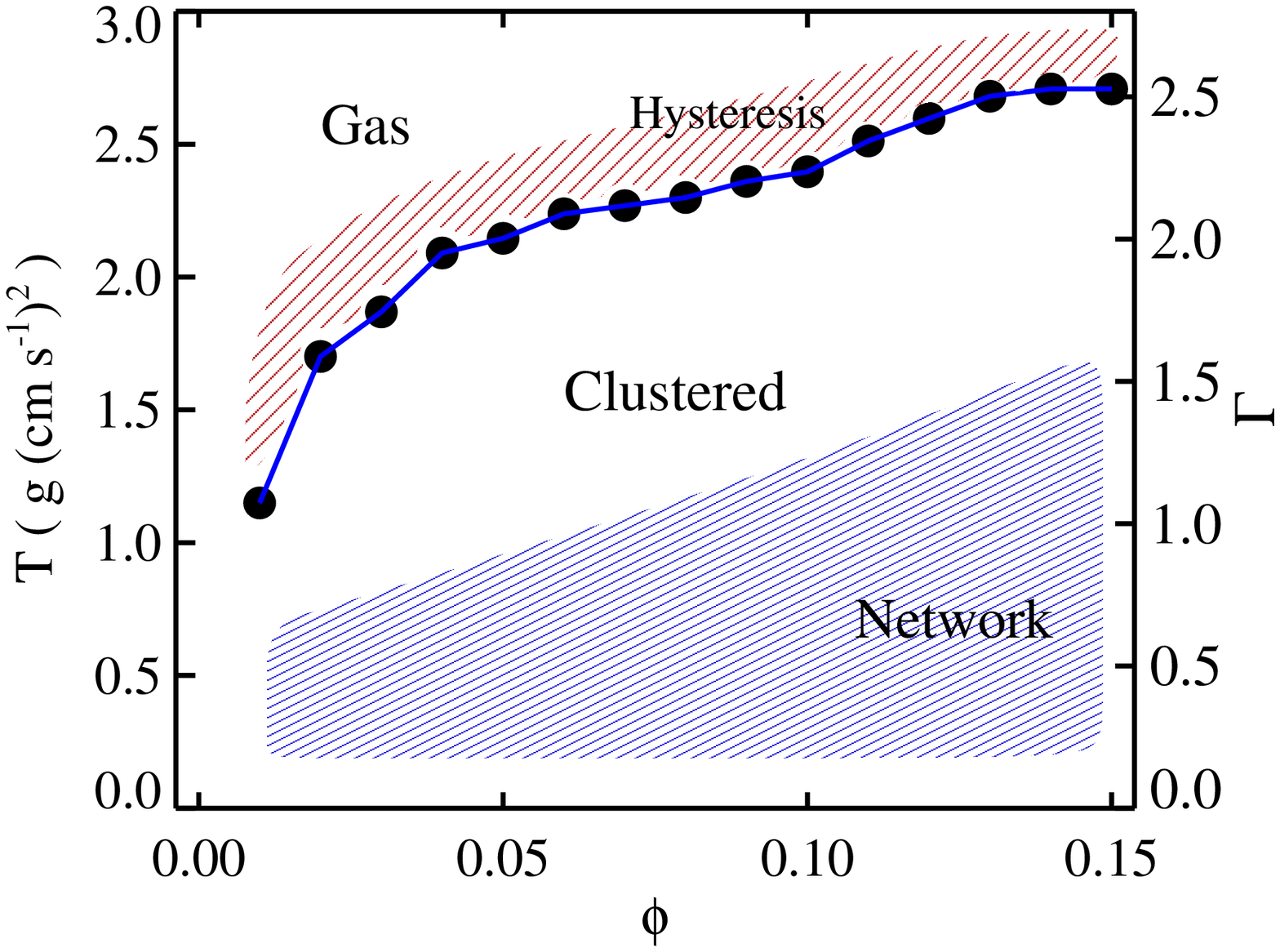,width=7.0cm}}
\caption{(a) The phase diagram of temperature $T$ versus the covering
fraction of the particles $\phi$.  The plate acceleration $\Gamma$,
scaled by gravity, is also shown for clarity. A gas phase consisting
of single particles and short lived dimers and trimers are observed
above a transition temperature $T_s$ that depends on $\phi$, shown by
the solid points. To evaporate a cluster in the gas phase one must go
past $T_s$ denoted by the hysteresis region.  Below $T_s$, dimers and
trimers act as seeds to the formation of compact clusters that coexist
with single particles [Fig.~\ref{fig:images}(d)].  If $T$ is rapidly
quenched from the gas region to very low $T$ highly ramified networks
of particles form [Fig.~\ref{fig:images}(e)].}
\label{fig:vel_dists}
\end{figure}

Our experiments begin by maintaining a gas of magnetic and glass
particles at a particular $\phi$.  By lowering $T$, a fraction of the
magnetic particles begin to self-assemble into the simple structures
shown in Fig.~\ref{fig:images}(a--c).  We define the temperature at
which this occurs as the transition temperature, $T_s$.  These
structures are transient, and over time will evolve into large
clusters that coexist in steady state with single particles.  An
example where a single ring has grown in time to form an extended
cluster, is shown in Fig.~\ref{fig:images}(d)~\cite{movie}.  However,
clusters of a very different conformation are produced if $T$ is
quenched far below $T_s$.  Figure~\ref{fig:images}(e) is an example of
the system after a rapid quench.  This extended mesh will eventually
rearrange to take on more a compact form, therefore we consider this
phase to be a meta-stable version of the clustered phase.

To characterize the gas and clustered phases observed, and the nature
of the transitions, we plot the phase diagram
[Fig.~\ref{fig:vel_dists}]~\cite{calibrate}.  The connected line shows
the transition temperature $T_s$ as a function of $\phi$. Above $T_s$,
short lived dimers and trimers are observed along with single
particles, but at and below this temperature, small seeds of magnetic
particles  [Fig.~\ref{fig:images}(a--c)], precipitate from the gas.
Initially seeds take on the form of short chains that quickly become
either rings or more compact states (it is noteworthy to mention that
rings $\ge 4$ particles are energetically more favorable than a chain
of equal particle number~\cite{degen_pinc,wen}).
Figure~\ref{fig:images}(d) shows an image of the system with $\phi
=0.09$ after 1092~s at $T_s$.  If $T$ is increased above $T_s$,
clusters show hysteretic behavior.  This hysteresis depends on the
ramping rate of $T$, and is not observed to disappear over laboratory
time-scales. Therefore it appears that the observed phase transition
is first order.

\begin{figure}[t]
\centerline{\epsfig{file=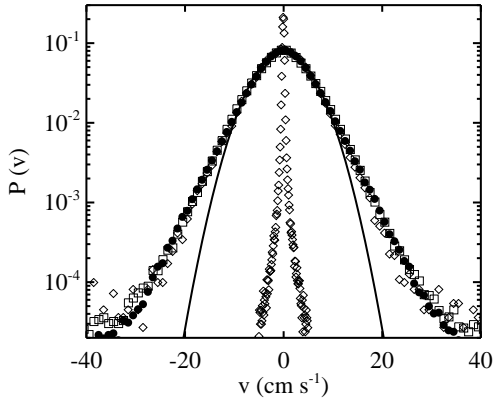,width=7.0cm}}
\caption{The probability distribution functions $P({\bf v})$ of the
velocity components in the gas and clustered phases on a log linear
scale. ($\diamond$) The p.d.f.'s for the magnetic particles in both
the gas and clustered phases.   The narrow distribution is from
particles within a cluster and the broad distribution (of the same
symbol) is for the surrounding gas.  ($\Box$) Magnetized, and
($\bullet$) glass particles in the gas phase.  The velocities are
absolute velocities given in (cm s$^{-1}$) not rescaled by the
temperature.  The solid line is a Gaussian fit to the magnetized gas
distribution.}
\label{fig:temps}
\end{figure}

In the gas phase, we measure both the velocities of the non-magnetized
glass particles and the magnetized steel particles and find that their
distributions are indistinguishable. Figure~\ref{fig:temps} shows the
distributions of velocities at $\phi =0.09$ for each particle species,
viz. the gas of glass particles, the gas of magnetized steel particles
(prior to and after nucleation of a cluster), and the particles within
a cluster.  The data shown corresponds to the system in
Fig.~\ref{fig:images}(d), although the results are the same for
different values of $\phi$. When particles nucleate into clusters the
fluctuations in the particle positions compared to the surrounding gas
dramatically decreases, as shown by the narrow distribution in
Fig.~\ref{fig:temps}.   However, the surrounding particles still have
a temperature identical to that of the pure gas phase.  The ratio of
the system temperature to the cluster temperature $T_c$ is $T/T_c
\approx 60$.  This effect is inherent to the non-equilibrium nature of
this system, where effects due to friction limit the {\em thermal}
agitations expected in the constituent particles that comprise gels
and polymers~\cite{gels}.  The existence of a finite measurable
temperature in the clustered phase does imply however that particles
are still mobilized, which can be seen by the substantial
re-arrangements that occur during cluster growth.

\begin{figure}[t]
\centerline{\epsfig{file=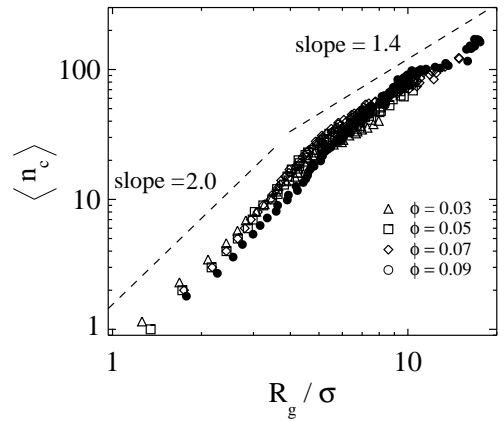,width=7.0cm}}
\caption{Average cluster size $\langle n_c \rangle$  versus the radius
of gyration $R_g/\sigma$.  The scaling $\langle n_c\rangle = \alpha
R_{g}^d$, where $\alpha = \pi^2 /2\sqrt{3}$, shows a crossover at $R_g
\sim 5 \,\sigma$~ indicating that the clusters start more compact and
become more extended.}
\label{fig:rg_gr}
\end{figure}

By utilizing the Hoshen-Kopelmen \cite{hoshen} cluster identification
algorithm individual clusters that form at $T_s$ are identified.  We
are then able to track each clusters' size, center of mass, and radius of
gyration. This method was also employed to measure the temperature of
a single cluster and the surrounding gas.  The average number of
particles contained in a cluster $\langle n_c \rangle = n_c/n$, where
$n$ is the total number of clusters formed, is plotted versus the
radius of gyration $R_g$ [Fig.~\ref{fig:rg_gr}]. $n$ is unity for most
times, but may vary at very short times. This data was obtained for
1092~s immediately after $T$ was lowered to $T_s$.  Using the scaling
relation $n_c \sim R_{g}^{d}$~\cite{scale}, we measure the compactness
of the clusters.  Figure~\ref{fig:rg_gr} clearly shows a crossover in
the dimensionality of the clusters.  Clusters with $R_g < 5.0\,\sigma$
follow closely to the spatial dimension.  As the number of particles
per cluster increases with time the trend for the fractal dimension,
$d$, at all $\phi$, is more consistent with $d = 1.4 \pm 0.1$.  This
implies that clusters are initially more compact and become  more
extended as they grow.   The change in dimensionality can be
understood by observing the growth of a cluster.  As  particles join
the cluster they do so by forming highly mobile chains that are
tethered by one end to the surface of the cluster. If the free end of
the chain is able to connect back onto the cluster an excluded area is
established.  During this process free particles may become trapped by
the chains,  which in time can produce rearrangements and re-opening
of these regions. This can be seen by a close inspection of
Fig.~\ref{fig:images}(d) and in Ref.~\cite{movie}.

\begin{figure}[t]
\centerline{\epsfig{file=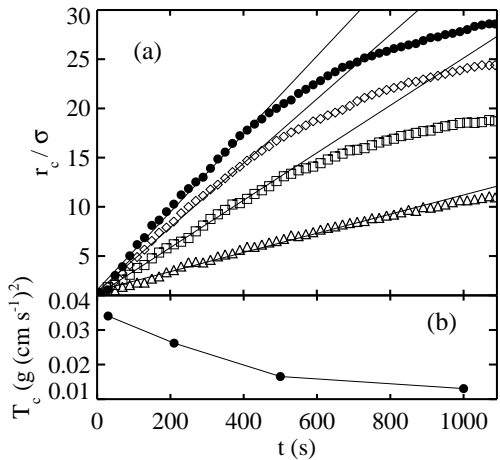,width=7.0cm}}
\caption{ (a) The radial growth rate $r_c/\sigma$ at $T=T_s$ for each
$\phi$.  The solid lines are fits to the data over a time when the
cluster is not in contact with the wall.  The apparent crossover in
growth rate is due to a finite size effect.  Linear growth is
consistent with a Wilson-Frenkel growth process, where individual hard
spheres are absorbed onto the  cluster. Symbols for each $\phi=$  0.03
($\+ \triangle$), 0.05 ($\+ \Box$), 0.07 ($\+ \diamond$), and 0.09
($\+ \bullet$). (b) The cluster temperature $T_c$ as a function of
time for $\phi = 0.09$.  The temperature of the surrounding gas
remains at a constant value of $T= T_s= 2.2$ g (cm s$^{-1}$)$^2$ }
\label{fig:time_dep}
\end{figure}

Next we plot the cluster growth rates by measuring the number of free
particles at $T_s$ for various $\phi$.  We find that the average
radius $r_c$ given by $n_c^{1/d}$ ($n=1$ for Fig.~\ref{fig:time_dep}
over all time) contained in a cluster grows linearly as the
surrounding free particles are depleted to a saturation value
determined by $\phi$.  The initial linear growth of $r_c$ is
consistent with the Wilson-Frenkel growth process of hard sphere
colloidal crystals, where individual particles are absorbed onto the
surface of a crystallite~\cite{ackerson95}, although the mechanism is
quite different.  The change in the linear growth of $r_c$ to a more
arrested rate is caused by two effects. As the clusters grow the
number of free particles is reduced and therefore the growth rate must
decrease as the system approaches steady state.  Data sets shown in
Fig.~\ref{fig:time_dep}(a) correspond to clusters that nucleate close
to the center of the cell.  Clusters are mobile and eventually drift
to the edge of the cell [see Fig~\ref{fig:images}(d)].   Due to
depletion forces~\cite{deplete}, the cluster will never be able to
return to the center of the cell once in contact with the sidewall.
As a result, half of the surface is not absorbing free particles,
therefore the growth rate is further decreased.

To characterize the temperature of each cluster and emphasize its
difference from the system temperature, we measure the cluster
temperature $T_c$ as a function of time at $\phi=0.09$
[Fig.~\ref{fig:time_dep}(b)].  $T_c$ is not only observed to be much
lower than the surrounding gas phase, but also observed to decrease
with increasing cluster size.

Finally, we discuss the meta-stable network phase that is observed
within the clustered phase, shown by the  shaded region of the phase
diagram [Fig~\ref{fig:vel_dists}].  This phase [see
Figure~\ref{fig:images}(e)], directly demonstrates the inherent
anisotropy of the potential described in Eq.~\ref{eq:dhs}.  This phase
is attained by quenching to a $T$ far below $T_s$ from the gas state.
When the energy of the system is dramatically decreased in this
manner, the particles quickly form chain configurations because
locally the head-to-tail alignment is energetically most favorable.
However, as discussed above,  when the length of the chain is $\ge$ 4
particles, the energy of the chain can be further lowered by bending
into a more compact structure.  At low temperatures the probability of
overcoming the potential barrier of bending is decreased, therefore we
are able to observe the meta-stable state that is comprised mostly of
string like objects for a longer period of time.  It is
self-consistent that the deeper the quench, the longer the network
phase survives, which is compounded even further by the dissipative
interactions between particles.  However, it is interesting that  this
phase shows similarities to recent Monte-Carlo simulations of
quasi-2D dipolar hard spheres~\cite{q2dhs}, where dissipation is not
considered.

In summary, we visualize the nucleation of clusters in a
non-equilibrium system by using magnetized granular materials.  We
have directly investigated the nature of the clusters formed by
measuring their temperature, dimensionality and their growth rates.
The temperature of the clusters is found to be considerably lower than
the gas phase indicating a breakdown of equipartition. We also find that
the phenomenology of the clusters is similar to that shown by recent 
work on dipolar hard sphere models, including a networked
phase~\cite{tlusty00,q2dhs}. Although our system is weakly
dissipative, it may provide still useful insights on the nature of
phase-transitions in ideal dipolar system due to our ability to
directly visualize and tune the particles in the experiments.

We thank C. Landee, H. Gould, G. Johnson, L. Colonna-Romano and
T. Tlusty for fruitful discussions. This work was partially supported
by National Science Foundation  under Grant \# DMR-9983659 and Alfred
P. Sloan Foundation.


\begin{thebibliography}{}
\bibitem{jaeger96} H. Jaeger, S. Nagel, and R.~P. Behringer,
Rev. Mod. Phys. {\bf 68}, 1250 (1996).

\bibitem{haff-jenkins83} J. T. Jenkins and S. B. Savage, J. Fluid
Mech.  {\bf 130}, 187 (1983); T.~P.~C. van Noije and M.~H.~Ernst,
Granular Matter {\bf 1}, 57 (1998).

\bibitem{simulations}C. Bizon, {\em et. al}, Phys. Rev. E {\bf 60}, 4340
(1999); E.~L. Grossman, T. Zhou and E. Ben-Naim, {\em ibid} {\bf 55},
4200 (1997); A. Baldassarri, {\em et. al}, {\em ibid} {\bf 64}, 011301
(2001).

\bibitem{los-olafsen99} S. Warr, J. M. Huntley, and G. T. H. Jacques,
Phys. Rev. E {\bf 52}, 5583 (1995); W. Losert, D. Copper, J. Delour,
A. Kudrolli, and J.  P. Gollub, Chaos {\bf 9}, 682 (1999);
J. S. Olafsen and J. S. Urbach, Phys. Rev. E {\bf60}, R2468 (1999).

\bibitem{krb} A. Kudrolli and J. Henry, Phys. Rev. E {\bf 62}, R1489
(2000);  D. L. Blair and A. Kudrolli, Phys. Rev. E {\bf 64}, 050301
(2001); K. Feitosa and N. Menon, Phys. Rev. Lett. {\bf 88}, 198301
(2002).

\bibitem{cohesive} P. Tegzes, {\em et al.},  Phys. Rev. E {\bf 60},
5823 (1999); A. Samadani and A. Kudrolli, Phys. Rev. E {\bf 64},
051301 (2001); I. S. Aranson, {\em et. al}, Phys. Rev. Lett. {\bf 84},
3306 (2000).


\bibitem{leeuwen93} M. E. van Leeuwen and B. Smit, Phys. Rev. Lett.
{\bf 71}, 3991 (1993).

\bibitem{sear96} R.P. Sear, Phys. Rev. Lett. {\bf 76}, 2310 (1996).

\bibitem{camp00} P. J. Camp, J. C. Shelley, and G. N. Patey,
Phys. Rev. Lett. {\bf 84}, 115 (2000).

\bibitem{tlusty00} T. Tlusty and S. A. Safran, Science {\bf 290}, 1328

(2000).

\bibitem{rosensweig} R. E. Rosensweig, {\em Ferrohydrodynamics}
(Cambridge University Press, Cambridge, 1985).

\bibitem{degen_pinc} P.~G.~deGennes and  P.~A. Pincus,
Phys. Kondens. Mater. {\bf 11}, 189 (1970).

\bibitem{mamiya00} H. Mamiya, I. Nakatani, and T. Furubayashi,
Phys. Rev. Lett. {\bf 84}, 6106 (2000).

\bibitem{wen} W.~Wen, F.~Kun, K.~F. P\'al, D.~W. Zheng, and K.~N. Tu,
Phys. Rev. E {\bf 59}, R4758 (1999).

\bibitem{movie} For a movie of this process please visit
\texttt{http://physics.clarku.edu/dipolar}

\bibitem{calibrate} We have performed a calibration between the
granular temperature $T$ and the  acceleration amplitue $\Gamma$.  We
find a nearly linear correspondence that does not depend sensitively
on the covering fraction $\phi$.

\bibitem{gels} A. G. Yodh {\em et al}, Phil. Trans. Roy. Soc. London A
{\bf 359} 921, (2001)

\bibitem{hoshen} J. Hoshen and R. Kopelman, Phys. Rev. B {\bf 14},
3438 (1976).

\bibitem{scale} $n_c = \alpha R_{g}^d$, where $\alpha$ is determined
by the dimensionality and mass elements.  We are 2D and use spheres
(disks), therefore $\alpha = \pi^2 /2\sqrt{3}$.

\bibitem{ackerson95} B.~J.~Ackerson and K.~Sch{\"a}tzel, Phys. Rev. E
{\bf 52}, 6448 (1995).

\bibitem{deplete} P.~D. Kaplan, J.~L. Rouke, A.~G. Yodh, and
D.~J. Pine, Phys. Rev. Lett. {\bf 72}, 582 (1994).

\bibitem{q2dhs}J. M. Tavares, J. J. Weis, and M. M. Telo da Gama,
Phys. Rev. E {\bf 65}, 061201 (2002).


\end{thebibliography}
\end{document}